\documentclass[pra,preprint,superscriptaddress,showpacs]{revtex4}
\usepackage{graphicx}
\usepackage{natbib}
\usepackage{latexsym}

\begin{document}



\title{A design outline for a Cherenkoff neutrino observatory}
\author{E.P. Bonvin \footnote[1]{Now at: Queens University at
Kingston, Sudbury Canada} }   \author{S. T. Hatamian
\footnote[2]{Now at: Mathematicus Laboratories, Sound Beach, NY,
email:tsh@MathematicusLabs.com}} \affiliation{California Institute
of Technology Pasadena CA 91125     USA }
\date{December-1991}

\begin{abstract}
An economical procedure to upgrade the existing Fairport water
Cherenkoff detector is described. The detector will lower its
energy threshold by increased photocoverage and light yield.
Furthermore, by addition of a Gd salt it will become sensitive to
neutrons produced by the inverse beta decay of anti-neutrinos on
protons. The new detector can then take advantage of the existence
of the Perry power nuclear reactor, located 12.9km away, as a
large source of anti-neutrinos. The present gap in the exclusion
plot of the neutrino oscillation parameters ($\Delta m^2$ from
$2\times 10^{-4}$ to $10^{-2} eV^2$) may be explored and closed
within 12-18 months of run-time. The detector will be able to
observe other types of neutrino sources (Boron-8 solar,
atmospheric, and supernova) with unprecedented advantages. The low
threshold also will allow searching for exotic modes of
proton-decay.
\end{abstract}

\maketitle
\section{Introduction}

The physics of neutrinos has been of increasing import among the
high energy community. This has been especially true in the last
ten years due in main part to the so called solar-neutrino puzzle:
the measurement\cite{Davis} in which the number of observed solar
neutrinos is significantly lower than that expected from the
"standard solar model". This has given rise to a surge of interest
in the neutrino-oscillation hypothesis. The oscillation hypothesis
suggests that the weak interaction neutrino eigenstates are a
superposition of the mass eigenstates, which therefore, whence
massive, may oscillate into one another. This oscillation may in
turn be responsible for the apparent deficit in the expected
electron neutrino flux, during its traversal of solar matter or in
the vacuum.

Terrestrial oscillation experiments have been carried out with
reactor neutrinos, accelerator generated (muon-)neutrino beams,
cosmic ray neutrinos, and atmospheric neutrinos. The upper limits
for the oscillation parameters are summarized in fig-1.
Oscillation experiments are usually analyzed in terms of a
two-parameter model (assuming only two neutrino flavors)
characterized by the mass parameter $\Delta m ^2 =  \mid m_2^2 -
m_1^2\mid$  line, and the mixing strength parameterized by an
angle: $\sin^2{2\theta}$ . The probability that a neutrino $\nu_l$
of energy $E_{\nu}$ will change into  $\nu_m$ after travelling a
distance $L$ (in vacuum) is given by:
\begin{equation}
 P_{\nu_{ml}} = \sin^2 {2\theta} \sin^2  {\frac{1.27  \Delta   m^2  [eV^2]  L[m]}{E_\nu [MeV]}}.
\end{equation}
While, the probability that $\nu_l$ will remain unchanged is given
by $P_{\nu_{ll}}=1-P_{\nu_{ml}}$.

In order to explore smaller mixing angles($\sin^2{2\theta}$) one
must increase the count rates of any given oscillation experiment.
To reach smaller mass parameters($\Delta   m^2$), however, one
must either search for longer neutrino oscillation wavelengths
(implies longer separation from the source, hence less
$\nu$-flux), or carry out the search for lower and lower energy
neutrinos($E_\nu$), or both. Particularly for the latter approach,
the low energy reactor neutrinos are the most attractive choice,
in addition to their relatively large flux, and an unrivaled
understanding and control of their source. In order to cover the
gap in mass parameter of the exclusion plot(fig-1), and given the
energy spectrum of a typical power reactor, the separation of the
detector from the reactor is then fixed to a value just over 10
Kilometers.

The Fairport facility (formerly used as a proton-decay detector)
is located 12.9km away from the Perry (3.8 GWatt) power reactor.
Though not designed as a neutrino detector, this fortunate
"accident" makes the Fairport facility a prime candidate for an
inexpensive and powerful neutrino detector. In this article we
delineate a method for upgrading of the Fairport detector to, not
only take advantage of the large source of neutrinos at the
perfect distance, but to in fact, serve as a neutrino observatory
to explore a host of neutrino related questions. These include
detection of not only reactor neutrinos, but also Solar,
atmospheric, and supernovae neutrinos. Moreover, the design can
make possible searching for proton decay via supersymmetric or
other exotic modes. Since the main point of the upgrade is a
lowering of the threshold of the detector, possible future
experiments with (Fermilab) accelerator neutrino beams are not
precluded.

Another proposal\cite{Steinberg} has sought to take advantage of
the proximity of the Perry nuclear power reactor, but using liquid
scintillator as the light producing medium. Though the light
produced by all forms of energetic particles is enhanced over
50-fold, many technical problems (relating to cost, handling,
containment, ppt-purification, and Gd-loading of the liquid
scintillator), make this alternative approach a challenge quite
different from the design obstacles tackled in the water detector,
described here.

In 1991, the IMB detector was decommissioned due to a leak which
resulted in the loss of much of the water, and some structural
damage. The discussion of this needed repair (albeit quite major),
as well as other necessary civil engineering considerations are
beyond the scope of our detector design synopsis, and will not be
addressed here. Here we will only assume that the structure can be
restored to a similar state as the original. In fact the necessary
reconstruction is an opportunity to re-build a much better
facility using the 12-years of operational experience, and indeed
better suited to the new design. Alternatively, the design and the
accumulated expertise may be transported to a whole new location
at the proper distance from a nuclear reactor in a new
underground, or submontane facility.

\section{An Outline And The Physics Focus}
The Fairport detector is a large tank which contains a total mass
of 8Ktons of water, purified through reverse osmosis. The present
fiducial mass is 3.3Ktons at a depth 1570 meters of water
equivalent(mwe). It is located at a Morton-Thiokol corp. salt mine
in Fairport, Ohio($41.7^o N, 81.3^o W$). The tank is 18m high, 17m
on the N-S side and 22.5m along the E-W direction. Presently, it
is designed to hold 2048 photomultipliers(PMT) of 8-inch diameter
to give a photocoverage of approximately 6\% of the walls, for an
effective threshold of just under 20MeV of electron energy.

Our proposal will make the primary mode of detection, the inverse
beta decay of anti-neutrinos on protons in the water. This
reaction produces a positron and a neutron. After thermalization,
the neutron via a large capture cross section on a nucleus such as
Gd-157 (=240Kb) is converted into gamma rays, and hence becomes
detectable. The threshold for a clean trigger must be reduced to a
few Mev gamma rays from the present 20 MeV recoil electrons. This
enables the detector to see reactor anti-neutrinos whose spectrum
ranges from 0 to 8 MeV.

The low energy of the neutrinos observed will allow limits on
smaller mass parameters on the mixing exclusion plot. The high
flux of the reactor is another attractive feature: at $2\times
10^7 /cm^2s$ (at 12.9km) this is larger than any atmospheric,
accelerator or solar source. The backgrounds in this mode of
anti-neutrino detection are made very small by requiring a
coincidence between the positron and the neutron signal, as the
definitive signature of the anti-neutrino. Moreover, the ability
to turn the source off and on, affords an unparalleled advantage
in measuring backgrounds. In this manner, the proposed upgrade
should inexpensively and speedily cover the exclusion plot gap
between the mass parameter of $2\times 10^{-2} eV^2$  to $10^{-4}
eV^2$ . If the parameters of the present proposal are successfully
implemented, the expected rate of detected neutrinos will be
around 10/day. \emph{This would allow to close the gap in a matter
of a year's running time}. Background (reactor off) and systematic
runs should take of order months. The detector may concurrently,
as well as later on, observe neutrinos from other sources with
much advantage over other techniques (presently in operation), as
will be discussed later in this paper.

\section{Discussion Of The Upgrade Task}
The upgrade project may be summarized in a list of 3 items:
        1. Create sensitivity to neutrons.
        2. Lower the detector threshold to 5 Mev.
        3. Reduce sources of potential backgrounds in this energy regime.
Each of these tasks will be discussed in detail in this section.

\subsection{Neutron Sensitivity}
\subsubsection{The choice of target nucleus}
The best candidate for detection of neutrons is the gadolinium(Gd)
nucleus. The first use of Gd additives as a detecting agent
(particularly in liquid scintillators) dates back to early 1950's
(see for example Ref-7 for an early mention in the context of
large detectors). Gd is a naturally occurring element comprised
of(15.7\%) Gd-157 and (14.7\%) Gd-155 (among its other isotopes).
The average thermal neutron capture cross section for natural Gd
is 49Kb. The radiative capture reaction will release approximately
8Mev of energy in an average of 3.5 gamma rays with one quantum
guaranteed to have at least 5.6Mev of energy. Cherenkoff light is
then produced by electrons from the Compton scattering of these
gamma rays. The feasibility of a similar process has been
demonstrated by Kamiokande-II \cite{Nakahata} who used a Cf-252
source as a neutron emitter embedded in a block of Nickel as the
capturer. Of the 9MeV of released gamma rays, they observed 7.8MeV
deposited well above backgrounds, which also agreed well with
their Monte-Carlo simulations. The actual case of Gd was simulated
by researchers in Sudbury Neutrino Observatory(SNO)\cite{SNO-prop}
Their results shows that nearly 60\% of the Gd gamma rays would be
detectable with a threshold of 4.5Mev at(40\% of 4$\pi$ -steradian
light-coverage). These studies suggest that as a rule, in this
energy range, gamma rays convert some 85\% of their energy into
electron energy (Cherenkoff light). Based on SNO's study, the
efficiency of neutron detection is plotted as a function of
detector threshold (or photocoverage), in fig-2. A threshold of
approximately 5MeV at 44\% photocoverage seems a reasonable goal.
The above value for photocoverage is not unreasonable:
Kamiokande-II already has a 20\% photocoverage and a threshold of
about 8MeV or less, but they are not fortunate enough to be near a
nuclear power reactor (see also photocoverage below).

Further candidates (see table-1) for the neutron capture target
may be natural Cd which yields slightly over 9MeV of gamma rays,
but at 2.5Kb cross section, one would need 20 times more Cd nuclei
to match the effect of Gd. Large concentration of additives might
reduce the light transmission of the water. Another candidate is
natural Sm which has an average cross section of 5.8Kb, and 8MeV
of gamma rays, in return its fast neutron capture cross section
might help with the timing of positron coincidence and thus
increase signal to noise.

 \vspace{10 pt}
\begin{tabular}{|c|c|c|c|c|c|c|}
  \hline
  Target & Mass(amu)  & $\sigma_{thermal}$(barn)  & Concentr.(wt.\%) & N($10^{20}/cm^3$) & Capt.Time($\mu S$) & $\gamma$ signal (Mev)\\
  \hline
  H  & 1.008  & 0.33  &  15 &  773  &  177  &  2.2 \\
  Cl & 35.45  & 33.5  &  20 &  34.3 &  39.5 &  8.6  \\
  Cd & 112.4  & 2450  &  1  &  0.472 & 39.3 &  8.9  \\
  Sm & 150.4  & 5,800 &  0.42 &  0.150 &  43.6  &  8.0\\
  Gd & 157.3  & 49,000 & 0.05 &  0.017 &  46.0  &  8.0\\
  \hline
\end{tabular}
\vspace{10 pt}

\emph{ Table-1: Summary of candidate neutron capture nuclei and
their properties.} \vspace{30 pt}

\subsubsection{Containment of the Gd loaded water}
Due to the need to continuously filter the water that is in
contact with the electronics and cables, one may not allow the Gd
loaded fiducial water to mix with the non-fiducial water. This
necessitates the need for a transparent containment vessel or bag
in which the water-Gd solution is contained, and is never routed
to the reverse osmosis filters. These filters will very
effectively extract the Gd-salt dissolved in the water. Fig-3 is a
sketch of the proposed scheme: A cube of 13m (2.2Kton) on a side
made of(1-2 cm thickness), transparent Teflon is suspended at the
center of the detector. This bag (or vesssel) will contain highly
purified(of pollutants and radioactivity) water, loaded with a
Gd-salt(and possibly wavelength shifters, discussed later). Note
that the transparent bag needs only to contain the fiducial mass.

In case of a breakage, all is not lost. The water may be purified
and the Gd is recovered, and the fiducial volume may be
reconstructed. Alternatively, one can consider a solid,
transparent vessel (e.g. of Acrylic) to contain the fiducial
water. The vessel material can potentially cause, a large loss of
light within it. For example, SNO's 5cm thick Acrylic vessel is
meant to securely contain the very expensive heavy water, but it
is responsible for a 21\% loss of light\cite{SNO-prop}. Light loss
can only be compensated by larger, costly photocathode area, or
smaller detector volume (closer-in walls). Another consideration
is the radiopurity of the bag or vessel material, This can become
a major contributor in the case of acrylic\cite{SNO-prop}. A thin
Teflon layer however can be made at significantly lower levels of
U and Th and contamination.

Teflon can also can help to bring the fiducial mass (with the
added weight of Gd-salt) to neutral buoyancy. It is, furthermore,
a good choice due to its physical stability, and lack of chemical
interaction with other material. However, careful studies of the
growth of bacteria on the Teflon are of paramount import in the
present design. The growth of such bacteria on glass and plastic
media has been observed, and has proven quite a nuisance despite
countermeasures in the water purification process. Light loss due
to bacteria growth is a universal problem which all large
water-based detectors such as SNO and Kamiokande must contend
with. In our case, the presence of Gd salt(s), and possibly the
complex molecules of wavelength shifters (if used) may
significantly affect the growth rate of these parasites. The
complete lack of fresh air flow in the sealed bag may prove a
favorable factor.

\subsubsection{Photocoverage}
In order to lower the threshold of the detector, more light
coverage is necessary. For a  5MeV threshold, this amounts to an
effective 44\% photocoverage. This may be achieved by simply
increasing the area of the photocathode coverage. Alternatively,
with a reasonable design for light-collecting reflectors, it seems
feasible to lower the photocathode coverage requirement to below
35\%. These figure are not unreasonable: we note that SNO and
SuperKamioka proposals call for similar or even larger figures
compared to 44\%.

The PMT's are mounted on a rigid outer shell, moved inward 2m from
the walls to increase photocoverage, as well as to allow a buffer
region to shield from low-energy radioactivity of the rock, salt,
etc. of the walls. Another purpose of the 2m region is to install
a small number of the PMT's, facing out, to act as veto triggers.
Another 1m is allowed in between the containment bag (fiducial
volume) and the PMT structure, to shield against radioactivity
from the glass in the PMT's, other electronics, and objects on the
mounting structure. The PMT's will need to be magnetically
shielded. However, a mu-metal grid for instance, can cut the light
transmission by up to 10\%\cite{Nakahata} This spells out further
need for added light collection by a reflector around the PMT's.

\subsubsection{On Adding a Wavelength Shifter to Increase Light-yield}
Since the bulk of the cost of the proposed upgrade is expected to
be that of the PMT price tag, further reduction in the required
photocathode coverage is desirable. One method is suggested by
noting that the Cherenkoff light spectrum has a significant
component in the UV\cite{Porter}, a region that is not detected by
ordinary photocathodes. Water soluble chemical additives, called
wavelength shifters, exist whose molecules will absorb the higher
energy photons and re-emit lower energy(visible) photons. While
this process will increase the light collected from the Cherenkoff
process, its shifted component will be isotropic and will not have
the directionality of the Cherenkoff cone; this light may be used
in a calorimetric mode only. Unfortunately, some (~20\%) of the
visible(directional) light from the Cherenkoff process may also be
absorbed and re-emitted thus resulting in dilution of the original
directionality\cite{Kielczewska}. While This loss does not affect
the energy-threshold, it will dilute the ability to reconstruct
particle tracks. The latter becomes a concern, mainly in the
search for proton-decay events.

A candidate for this task is
beta-methylumbellapherone\cite{Porter}. This liquid will shift
light from the wavelength range of 250-350nm to 380-540nm, by
absorbing light in the former range and re-emitting in the latter
range. At a concentration of 50ppm, the average absorption
length($\alpha$) is 5cm for  $\lambda \in$[250,350]nm ($\alpha >$
20m for [400,700]nm). At this level, a factor of 1.74 in
light-yield has been shown to be gained from the addition of the
wavelength shifter\cite{Kielczewska}. Although the loss of
directionality of light is an important concern, the use of a
wavelength shifter could prove a major improvement in the
effective photocoverage of the detector. Perhaps the development
of a customized wavelength shifter may prove worthy of the
investment to address the problem of loss of directional portion
of the light for better track reconstruction. To summarize, a
factor of 1.74 improvement in the light yield, will reduce the
final photocathode-coverage needed to 25\% (and even lower with
light collectors), while retaining an effective photocoverage of
44\%.

\subsubsection{Detection of neutron-positron coincidence: Electronics}
The sequence of events following a neutrino interaction on a
proton is as follows: first a positron (0-5.5MeV, median 1.5MeV)
and a neutron are released. The positron will soon stop, giving a
weak Cherenkoff signal below threshold. Given the 0.05\%
concentration of Gd, by 100 microseconds later, the thermal
neutron is radiatively captured. The gamma cascade will last about
100 nSec. The mean neutron capture time is 46
microseconds\cite{Boehm}.

Although it is conceivable that the detection of the thermal
neutron's radiative capture will be sufficient to signify a
neutrino event, a much cleaner signal would be the observation of
the neutron-positron delayed coincidence. For the latter purpose,
one would require an on-line (hardware) trigger on all neutron
events, via the emission of nuclear gamma rays. Thereupon, the
output of a system of high rate waveform digitizers, with a memory
depth of approximately 100 microseconds, will be frozen for
offline(software) acceptance, or veto of the subsequent neutron
event. At the proposed levels of photocoverage, a neutrino event
will be accepted if a 20-25 photoelectron event was preceded by a
6-7 photoelectron burst in its immediate past 100 microseconds.
The accidental backgrounds for this event will be shown to be very
small indeed in the following section.

A digitizing system (8 bit dynamic range) with quite similar
characteristics as that needed here, has been demonstrated with a
memory depth of 500 microseconds operating at 500 MHz in
ref\cite{Atyia}.

\subsection{Background Considerations}
\subsubsection{Uncorrelated Background}
Uncorrelated backgrounds are comprised of the accidental
coincidences of relatively large singles rates. These are
estimated and listed in table-2, assuming a 2.2Kton fiducial
volume, and 5 MeV threshold. The sources labeled "internal" are
due to radioactivity within the fiducial volume and hence mostly
dependent on the purity of the water. The figures are scaled based
on SNO's careful computations at 0.02 ppt water (U and Th
content), and at 40\% photocoverage. In SNO's actual case, this
class of noise is some 10 times larger\cite{SNO-prop}, simply due
to the radioactivity in the acrylic containment vessel. A mass
spectroscopic analysis of the Fairport water yielded the U+Th
content of 20 ppt. Thus the fiducial water must be 1000 times
purer than the Cleveland city water before it is sealed in the
containment bag.

 \vspace{10 pt}
\begin{tabular}{|c|c|c|c|}
  \hline
  Type & Source  & Reaction  & Rate \\
  \hline
  INTERNAL & Water & Th+U at 0.02 ppt &10/day  \\
           & Water  & Spontaneous fission  & $\ll$ 1/day \\
           & PMT's  & ($\alpha$ ,$p \gamma$) at 30ppb U+Th  & 10/day \\
  \hline  EXTERNAL & Walls  & Muon spallation on NaCl  & 0.01/day \\
           & Walls  & ($\alpha$ ,p$\gamma$) on Na at 30ppm U+Th  & 3/day \\
           & Walls  & ($\alpha$ ,p$\gamma$) on Al in concrete & 5(x\%Al)/day \\
           & Walls  & ($\alpha$,n) and(n,$\gamma$) on Na  & 0.03/day  \\
   \hline COSMIC & Muons  & $\mu O^{16} \rightarrow \nu_\mu N^{16}$  &  30/day \\
         & Muons  & muon decay  &  2000/day \\
         & Muons  & spallations &  600/day \\
  \hline
\end{tabular}
\vspace{10 pt}

\emph{Table-2: Estimated rates of singles events in the proposed
detector. Figures are based on 2.2Kton fiducial water and 5MeV
threshold.}

\vspace{30 pt}

The "external" class of backgrounds is referred to all sources of
radioactivity which originate outside of the water. These sources
are negligibly small, thanks to the large buffer regions allowed
in between the fiducial volume and the walls.

Noise due to cosmic ray muons that interact directly with the
fiducial volume are by far the largest source of uncorrelated
background. These are primarily due to decay of muons inside the
fiducial volume. Muons also cause break-up of nuclei present in
the water. A small number of muons will convert O-16 into N-16
which will beta decay(Q=10.4 MeV) with a mean-life of 7-Seconds.
Muons may decay inside the fiducial volume at a large rate. The
released electron has an energy up to 100MeV. Muon induced
spallations are rarer but more complex, as a variety of beta decay
isotopes with $A<16$ may be created; they have lifetimes ranging
from 10-3 to 1 second, and $Q>10$MeV.

Given the rates above and the coincidence window of a
positron-neutron coincidence, the uncorrelated backgrounds pose
little concern for the detector's signal to noise. This is so,
even without the constraint that the neutron and the positron
events must appear within an approximately 8$m^3$  volume due to
finite travel time of the thermalized neutron.

\subsubsection{Correlated Background}
Of much greater concern, are correlated sources of noise:
Non-neutrino events that mimic a positron-neutron coincidence
signature. This is particularly problematic in the case of muon
spallations: often a neutron is knocked free in addition to the
creation of beta/neutron emitters(see table-3). As the result, the
two ingredients of a genuine neutrino event are mimicked. The
solution to this problem is the existence of the strong muon track
that precedes the electron-neutron event. In addition, if not
missed, the (one or more) freed spallation neutrons will precede
the beta event, which will make the event even more distinguished
as a spallation reaction.

 \vspace{20 pt}
\begin{tabular}{|c|c|c|c|}
  \hline
  Isotope & Lifetime  & $Q_\beta$  & Branching ratio \\
  \hline
  $Li^9$    & 0.2 sec  & 13 MeV  & 35\%  \\
  $Li^{11}$ & 10 msec  & 21 MeV  & 61\%  \\
  $C^{16}$  & 0.7 sec  &  4 MeV  & 100\%  \\
  \hline
\end{tabular}
\vspace{10 pt}

\emph{Table-3: Spallation isotopes that decay via prompt
beta-neutron emission. The expected rate of these events are
estimates around 35/day before rejection due to proximity to a
strong muon track. \vspace{30 pt}}

The detector's larger photocoverage should further pay off here:
the improved tracking of trajectories is needed to reject
spallation events. The precise rate of these muon related events
will also be measured during the reactor-off periods. In all, a
high signal to noise ratio is in principle, achievable for
detecting 10-15 reactor neutrinos per day.

\section{Non-Reactor Neutrino Physics}
A host of new neutrino physics questions may be explored using the
proposed low threshold water Cherenkoff detector. Perhaps the most
important is the case of the Boron-8 solar neutrinos. To date, the
foremost electron-neutrino scattering experiment has been that of
Kamiokande-II which has a threshold of around 8MeV. Fig-4 shows
the energy spectrum of these neutrinos as transformed into recoil
electrons. Clearly a threshold of 5MeV will vastly improve the
observation of the Boron-8 neutrino flux. Based on extrapolations
from Kamiokande rates, the proposed detector will observe over 6
events per day of Boron-8 neutrinos, some 20 times larger than the
present leader in this field, Kamiokande. For comparison, the
proposed Super-Kamiokande detector\cite{Kitamura} will expect to
see about 23 events per day.

Atmospheric neutrinos are another handle on possible oscillation
phenomena. The observation of relative deficiency of muon
neutrinos to electron neutrinos is fairly well established.
Similarly to the case of the Boron-8 spectrum (fig-4), the
upgraded Fairport detector with a low threshold, will improve
observed rates over those achieved by Kamiokande-II. Moreover,
there will be less difficulty with systematic errors in energy
calibration, as it easily avoids very steep parts of the
atmospheric neutrino energy spectra starting at just above 10MeV
of electron recoil energy. With improved rates and less systematic
effects, more precise comparisons of up-going vs. down-going
neutrinos may be made, to search for terrestrial MSW effect.

If to occur during the life of the upgraded detector, supernovae
neutrinos will yield unprecedented insight into mechanisms of
stellar explosions. This is achieved because both of the expected
neutrino and anti-neutrino bursts may be observed, and
distinguished from one another. Neutrinos will be observed via
electron scattering. The expected flux spectrum of each of these
bursts are plotted in fig-5. Note that during the 1987A
super-nova, Kamiokande (0.68Kton) had a threshold of about 8MeV
and the Fairport (IMB-) detector (3.3Kton) had a threshold of
about 20MeV. In addition, both could only observe the lower curve
of neutrino-electron scattering. Whether or not there are two
bursts of neutrinos, and how far apart they are in time, will
confirm or refute entire classes of supernova
theories\cite{Burrows}.

Finally, search for supersymmetric proton decay could continue. In
supersymmetric theories the proton may, for example, decay into a
neutrino and a kaon, or a pion. Previously, The neutrino-kaon
branch was searched for by looking for the muon decay of the kaon.
Many new lower energy decays may be sought with a lower threshold
detector. For example, one could consider the proposed
recombination of nucleon-hole pairs as put forward in
Ref\cite{Ejiri}. This mode may only be investigated with a
detector threshold of order 5MeV or lower.

\acknowledgements{Much of the work on the details of the present
proposal was done by E.Bonvin. I also wish to thankfully
acknowledge numerous useful discussions with the members of the
IMB group, especially: F.Reines, H. Sobel, W.Kropp, C.McGrew, And
P.Halverson. This work was supported by the US Department of
Energy.}

\newpage
FIGURE CAPTIONS

Fig-1. Exclusion limits for neutrino oscillation from experiments
to date. The lower hollow region is the expected range of
parameters for the MSW effect in Boron-8 solar neutrinos. The
dashed curve outlines the goal of the upgraded Fairport detector.

Fig-2. Plot of neutron capture efficiency as a function of
detector threshold, from ref.\cite{SNO-prop}.

Fig-3. Schematic of the basic design proposed. The fiducial volume
is defined by the transparent Teflon bag.

Fig-4. The recoil electron spectrum of solar Boron-8 spectrum.
Kamiokande-II has a threshold around 8 MeV. Note the extent of
improvement with a threshold at 5MeV and with three times the
fiducial volume.

Fig-5. The theoretical recoil electron spectrum of a
supernova\cite{Burrows}. Both Kamiokande and Fairport/IMB were
sensitive to the lower curve only, during the SN1987a.
Fairport/IMB had a threshold of around 20MeV.

\end{document}